# The Cornstarch Flamethrower

**Thomas Concannon**
thomasconcannon@kings.edu

**Abstract**

Igniting cornstarch powder is a classic physics demonstration that showcases the rapid conduction of heat for a material in which the surface area is greater than the volume of its constituent particles. Including such a demonstration in a physics "magic show" for the general public presents certain challenges such as reproducibility and consistent crowd appeal. A simple but effective design for widely scattering cornstarch dust over a flame breaches these challenges and always results in consistently large, crowd-pleasing fireballs; so much so that the resulting demonstration has been dubbed the "cornstarch flamethrower." A small-scale version may also be used effectively for classroom instruction.

**Key Words:** Cornstarch, Thermodynamics, Dust explosion, Demonstration

**PACS Numbers:** 01.50.My, 05.70.-a, 44.10.+i

## I. Introduction

Doing physics "magic shows" for the general public or for local area schools is usually an integral part of any physics department's outreach program. These demonstration shows should not only teach fundamental physics principles with "standard" demonstrations (like the rocket cart), but should also include the "wow!" types of demonstrations for maximum audience impact. Presenters quickly discover the audience's thirst for spectacular demonstrations; if too few "wow!" demonstrations are performed, the presenter can lose the audience's attention or worse, the audience may get bored and leave. Some of the more exciting demonstrations include, for example, the Van de Graff generator, a large Tesla coil, the bed of nails, liquid nitrogen experiments, imploding gas cans, breaking a glass container with sound waves, and igniting cornstarch powder [1]. The focus of this paper is a suggested improvement to this last demonstration, to enhance the "dust explosion" effect for demonstration show purposes. At the end, I also describe a small scale version for use in classrooms, where the physics and chemistry of dust explosions can be explored in more detail than possible in a demonstration show atmosphere.

## II. Physics Demonstration Show Challenges

While coordinating demonstration shows for the general public sponsored by the University of North Carolina at Chapel Hill physics department [2], I had the opportunity to experiment with most of the in-house demonstrations. I wanted to spruce up some of these demonstrations to increase the "wow!" effect. The goal was not only to make the demonstrations more dramatic overall but to solve the more difficult problem of obtaining *consistent* results. Too many times I had to relegate demonstrations to the "boring" category because they did not quite work the way they should when they should. For example, one of the more notoriously unreliable demonstrations was the Van de Graff generator; on humid days it did not build up enough static electric charge to spark consistently. Another unreliable demonstration was our version of the cornstarch dust explosion. Rather than the environment, the source of the unreliability was essentially in the equipment I used to showcase the explosion.

So my immediate goal was to create a cornstarch dust explosion that was suitable for a large audience during a physics demonstration show, one that was both reliable and one with a high "wow!" factor.

## III. Original and Traditional Cornstarch Dust Explosion Demonstrations

Our cornstarch explosion apparatus consisted of cornstarch loosely packed into a plastic drinking straw which was then blown over a candle flame. The intended result was a small fireball, ignited when the cornstarch powder was dispersed over the flame. While nicely adequate for classroom use, and to which we come later in this paper, it wasn't satisfactory for a large demonstration show. Sometimes the puff of air would be too strong and blow out the candle flame. The presenter would then need to pause to re-light the candle, re-load the straw, and try again. In a demonstration show setting, the audience could quickly grow impatient if the re-lighting/re-loading process occurred frequently.

At the time, the world wide web was in its infancy, so searching for alternative demonstrations was more difficult. Now it is easy to find that there are others who do the cornstarch dust explosion a little differently [3]. In one alternate design, for example, the cornstarch is placed in a metal bucket directly underneath an opening to which an external tube is attached. Opposite the cornstarch in the bucket, a candle flame is placed. When air is pumped into the bucket, it disperses the cornstarch powder over the flame, which ignites into a large flame erupting from the bucket (see Figure 1). This demo results in a nice "silo explosion" reminiscent of actual exploding corn silos due to improper ventilation and/or improper electrical wiring.

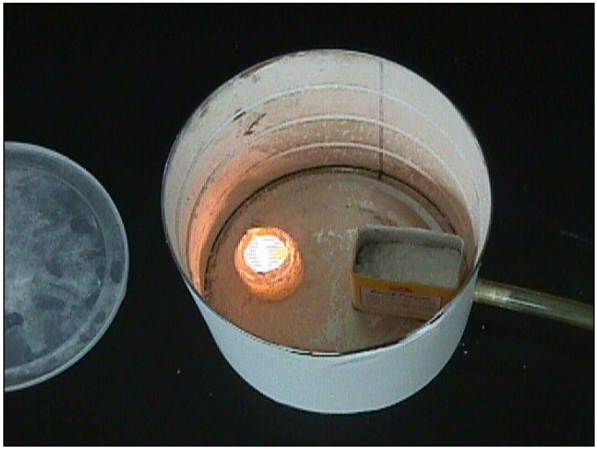

**Figure 1a: Cornstarch silo explosion setup**
(Picture used with permission [4].)

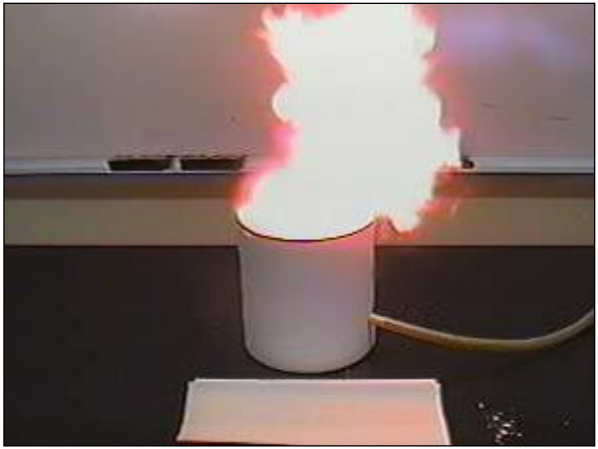

**Figure 1b: Cornstarch silo explosion**
(Picture used with permission [4].)

Though this more traditional demonstration (it is, in fact, the only dust explosion demonstration now listed on the PIRA web site [5]) is definitely reliable and dramatic, I only wanted to upscale our existing demonstration for large audience consumption and make it more consistent in producing fireballs. The smaller classroom-based design, therefore, needed some tweaking for use in a large demonstration show setting.

The first revision involved replacing the straw with another instrument of potential dispersion, a wax paper cup. I took a 12 ounce wax Dixie® cup and cut out a small semi-circular hole, approximately the size of half of a quarter, in the bottom. The hole was taped and the cup was filled with cornstarch. Then I cupped the drinking end to my mouth and blew the cornstarch over the candle flame. Two things went wrong with this approach: (1) I suffered from cornstarch powder "blow back" with it all over my face; and (2) my puff blowing the cornstarch was too strong and subsequently blew out the candle flame!

To counter the latter problem, I replaced the candle flame with a much larger flame: a Bunsen burner flame, whose fuel source is a small camping propane tank to allow for easy portability (see Figure 2).

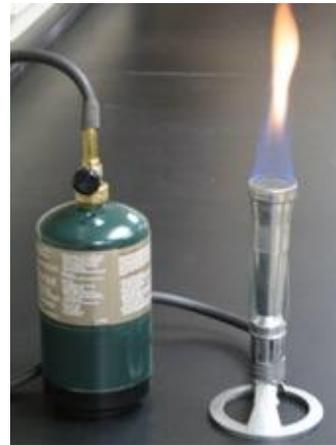

**Figure 2: The Bunsen burner flame and fuel source**

To alleviate the cornstarch blow back problem, I decided to take two of the Dixie® cups that had the semi-circular holes cut out of their bottoms and put tape over the holes so that they may be eventually transported without leaking. I filled one almost entirely with cornstarch, and then taped their tops together tightly with strong, transparent Scotch™ tape, their drinking ends now coincident (see Figure 3) and such that the holes were 180° apart relative to each other.

**IV. Improving the Demonstration**

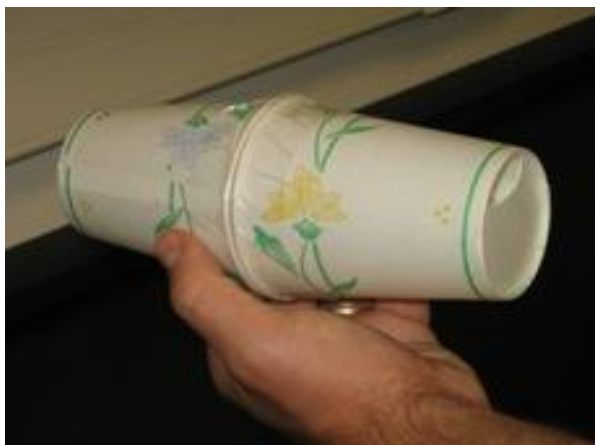

**Figure 3: The improved cornstarch dust dispenser**

## V. Performing the Demonstration

Before doing this demonstration, adequate safety precautions for both the audience and the presenter should definitely be followed. First, there should be enough space between the presenter and the audience, ideally up to 2 or 3 meters. The rule of thumb is that the audience should not feel the heat of the fireball at any time. Second, a fire extinguisher should also be close by in case of any accidental ignition. Third, the presenter should place the Bunsen burner on a flat surface, such as the lecture table at the front of the room, so that no one is then required to hold it. Fourth, clear the table of any flammable material whatsoever so that the Bunsen burner is the only object on the table. Fifth, the presenter should wear safety goggles to protect the eyes from any cornstarch powder blow back. Finally, I recommend using a small Plexiglas® shield, perhaps a circular one with about a 2 foot diameter, with a hole cut in the center into which one end of the device may be snugly inserted. Using this shield, the presenter is then protected from the heat of the resulting fireball.

The demonstration presenter should also practice a few times before the show begins. Practice is necessary to determine how much breath and what strength of puff to use to produce a desired fireball size. For large demonstration shows in a large lecture room, a deep breath and a strong puff will produce a fireball about a meter long and half a meter wide. For smaller rooms, a less deep breath but a similarly strong puff to spray the cornstarch over the flame in a fine powdery form will still be needed. In this case, a smaller, but still crowd-pleasing, fireball is produced. If it happens that the presenter accidentally doesn't blow hard enough through the dispenser, the cornstarch will not spray over the flame and the clumps of cornstarch that emerge from the dispenser will most likely just vanquish the Bunsen burner flame rather than produce any fire. The resulting fireball is literally over in a flash, with a lifetime on the order of about a second or less, depending on its size.

The dispensers should be assembled and filled with cornstarch before the show starts, in a location other than where the demonstration will actually take place. This way no stray cornstarch powder near the demonstration area can unexpectedly ignite. In fact, I usually prepare 3 to 4 dispensers before the show. It turns out that one strong puff is usually enough to empty the entire dispenser of cornstarch, and thus one entire dispenser equates to only one fireball. Since the audience almost always demands an encore, preparing multiple dispensers beforehand is a good idea.

Once the above safety precautions have been followed, the practice adequately performed, and the dispensers filled and ready to go, the demonstration may be performed. Explain how a pile of cornstarch is not usually flammable, but can be under the right circumstances. Briefly explain some of the underlying thermodynamics (see the following section) and proceed with the demonstration. Listen for the "oohs" and "aahs" afterward, along with the unanimous request for you to do the obligatory encore (as I mentioned above). Because of its high "wow!" factor and because it can reliably generate large fireballs with a strong puff over a nice sized distance, I've dubbed this particular version of the cornstarch explosion demonstration, "The Cornstarch Flamethower."

## VI. Uses in the Classroom

In the classroom, an instructor can take advantage of the small-scale version of this demonstration while discussing the first law of thermodynamics (see Figure 4). First, for example, the instructor can emphasize how the energy transfer rate into a substance depends on the surface area of the reactants, while the rate of temperature change depends on the volume of the reactants. With a pile of cornstarch powder sitting on a tabletop in the front of the classroom, the instructor can proceed to try to light it on fire with any flammable source. Thinking that cornstarch may be non flammable and inert, the students may not be expecting the resulting little fireball produced by the small-scale cornstarch flamethrower the instructor subsequently performs (with a fire extinguisher nearby, of course)!

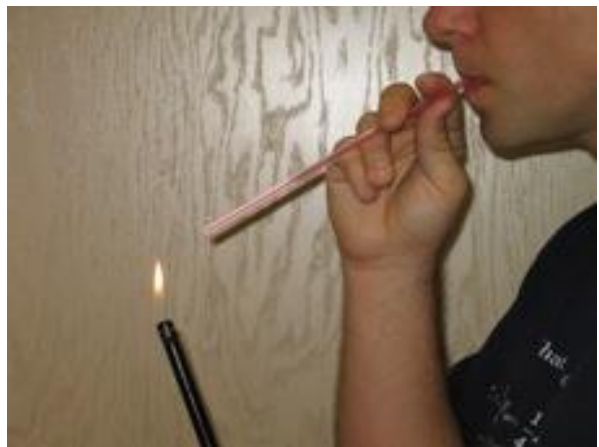

**Figure 4: The classroom version**

After the demonstration, the instructor can then lead the students through the differences between the two experimental setups. Since the cornstarch dust blown over the Bunsen burner flame has a larger surface to volume ratio per particle exposed to the heat than the pile of cornstarch does [6], the dust reaches a higher temperature more quickly due to the faster heat energy conduction rate from the combustion reaction. (Perhaps a more familiar example is that of heat conduction in metal— a thin wire will conduct heat energy faster than a thicker wire and thus will become hotter more quickly. A nice demonstration of this is trying to burn a steel rod, then actually succeeding in burning steel wool [7].)

The instructor can then lead the students into a discussion of what other factors effect the reaction rate, socratically obtaining answers such as the surface area of the reactants (which translates to the number of collisions per second), environment temperature, concentration of the reactants, environmental pressure, and the chemical nature of the reactants themselves. For this particular cornstarch combustion reaction, the reaction may be written as

$$(C_6H_{10}O_5)_n + 6nO_2 \rightarrow 6nCO_2 + 5nH_2O \quad (1)$$

where the **n** denotes the $n^{th}$ molecule in the cornstarch polymer. This can then lead into the physics of chemical reactions and the physics of combustion in general.

Finally, the class can discuss why this particular demonstration applies to real-world problems, such as coal mine explosions (of particular concern in my local area of northeastern Pennsylvania), corn storage silo explosions, and a seemingly more innocent example of blowing out the candles on a child's birthday cake, dangerously decorated with powdered sugar. Armed with such knowledge, students can aid in designing better and safer environments for future corporations, working individuals, and families, saving lives in the process.